\documentstyle[12pt,titlepage]{article}
\input epsf
\newcounter{subeqn}
\newcommand{\tr}{\, {\rm tr} \,}
\newcommand{\e}{\, {\rm e}}
\renewcommand{\P}{\Phi}
\newcommand{\p}{\phi}
\begin{document}
\begin{titlepage}
\addtolength{\baselineskip}{.7mm}
\thispagestyle{empty}
\renewcommand{\thefootnote}{}
\begin{flushright}
NBI-HE-96-01\\
hep-th/9601043\\
January 1996\\
revised February 1996\\
\end{flushright}
\vspace{10mm}
\begin{center}
{\large{\bf Wilsonian Approximated Renormalization Group\\
for Matrix and Vector Models in 2$<$d$<$4}}\\[15mm]
{\sc Shinsuke Nishigaki}
\footnote{\noindent e-mail address: nishigaki@nbi.dk} \\
\vspace{10mm}
The Niels Bohr Institute\\
Blegdamsvej 17, DK-2100 Copenhagen \O, Denmark\\[6mm]
\vspace{20mm}
{\bf Abstract}\\[5mm]
\end{center}
Wilson's approximation scheme of RG recursion formula
dropping momentum dependence of the propagators
is applied to large-$N$ vector and matrix models in 
dimensions $2<d<4$ by making use of their exact solutions 
in zero dimension.
In spite of apparent dependence of critical exponents
upon the dilatational parameter $\rho$
involved by the approximation, 
the exact exponents are reproduced for vector models
in the limit $\rho\rightarrow 0$.
Application to matrix models is then reexamined 
after the same fashion. 
It predicts critical exponents
$\nu=2/d$ and $\eta=2-d/2$ for the $\tr \Phi^4$ matrix model.\\

\end{titlepage}
\renewcommand{\thefootnote}{\fnsymbol{footnote}}
\setcounter{footnote}{0}

%%%%%%%%%%%% SECTION 1: INTRODUCTION %%%%%%%%%%%%
\centerline{{\bf 1. Introduction}}

The study of non-gaussian random matrix models
initiated by the pioneering papers
\cite{BIPZ} have been combined with Weingarten's idea of discretized
quantum gravity \cite{Wei} and yielded a thorough understanding of 
$c\leq 1$ noncritical strings \cite{DFGZJ}.
There the `$c=1$ barrier' \cite{Sei}, 
traditionally attributed to the tachyonic nature of
ground state of bosonic strings, obscures itself among  
the technical difficulty with the matrix model in dimensions $d>1$:
that it does not allow one to perform angular $U(N)$ integration 
so that the system is no more reduced into free fermionic eigenvalues 
confined in a potential well. 
Despite the importance of uncovering the nature of phase transition
of random surfaces around $c=1$,
several previous attempts proposed to circumvent this difficulty,
including the Br\'{e}zin--Zinn-Justin program \cite{BZJ,HINS} and
the light-cone quantization \cite{DK},
still fail to provide us with a reliable prediction 
even in the `planar' large-$N$ limit.

This letter is aimed to present an insight into this long-standing
problem, by viewing the $\tr \Phi^4$ matrix model as a Landau-Ginzburg 
hamiltonian and exploiting Wilson's treatment of 
the renormalization group \cite{W}. 
It consists of the following steps to derive an RG recursion formula:\\
\indent 
i) to separate $\Phi(x)$ into its high/low-frequency parts with respect 
to an arbitrarily 
\indent
chosen mass scale $\rho$\\
\indent 
ii) to perform functional integration over the high-frequency field by
using an approx-\\
\indent
imation of propagators
$1/(p^2+r)$ and $-d/dp^2|_{p=0}\rightarrow 1/({\rm const.}+r)$,
as well as trunca-\\
\indent
tion of induced interactions\\
\indent 
iii) to rescale coordinates and the low-frequency field to pull back
the renormalized\\
\indent 
action into the same form as the original action.\\
\noindent
The approximation ii) is equivalent to substituting 
all the loop integrations by
zero-dimensional combinatorics, which was readily put in our disposal by 
ref.\cite{BIPZ}.
This program was previously applied by Ferretti \cite{F} 
to a $d=3$ matrix model, although he relied upon an assumption of
universality for the choice of $\rho$ which proves incorrect in the sequel.

In this letter I first apply the program to $O(N)$-symmetric vector models
in the large-$N$ limit \cite{ZJ,HB}, 
which have served as a probe to matrix models due to 
their resemblance to and their simplicity relative to the latter
\cite{NY,HINS}.
I shall show an apparent $\rho$-dependence of the exponents
involved by the approximation, and then 
extract the known exact result
for $2<d<4$\footnote{
Due to the IR divergence this approximation is known to be 
inapplicable for $d\leq 2$.}
by taking the maximal dilatation limit $\rho\rightarrow 0$.
Next I reexamine the application to matrix models and
obtain the mass exponent $\nu=2/d$ and 
the anomalous dimension $\eta=2-d/2$
in the same limit.\\

%%%%%%%%%%%%%% SECTION 2: VECTOR MODEL %%%%%%%%%%%%%%
\centerline{{\bf 2. Wilsonian approximated RG for vector models}}

In this section we consider a Euclidean field theory of
an $N$-component scalar $\P(x)$ in dimensions $2<d<4$ 
with a cutoff, which we choose as the unit 
of mass. The action is
\begin{eqnarray}
S[\P]&=& \int {d^d x} \left[  \frac{1}{2} (\nabla \P)^2+\frac{r}{2}\Phi^2
+\frac{u}{N}(\Phi^2)^2 \right] 
\label{Svm}\\
&=&{1\over 2}\int_{0\leq |p| \leq 1} {{d^d p}\over{(2\pi)^d}}
(p^2 + r)\Phi(p)\cdot\Phi(-p)\nonumber\\
&+&{u\over N}\int_{0\leq |p_i| \leq 1}
{{d^d p_1\, d^d p_2\, d^d p_3}\over{(2\pi)^{3d}}}
\left(\Phi(p_1)\cdot\Phi(p_2)\right)
\left(\Phi(p_3)\cdot\Phi(-p_1-p_2-p_3)\right). \nonumber
\end{eqnarray}
Now we introduce an arbitrary mass scale $0<\rho<1$ and separate 
$\P$ into its low and high frequency parts accordingly, 
\begin{equation}
\P(x)=\int_{0\leq|p|\leq \rho}
\frac{d^d p}{(2\pi)^d}e^{ipx}\bar{\Phi}(p)+
\int_{\rho<|p|\leq 1}
\frac{d^d p}{(2\pi)^d}e^{ipx}\p(p).
\end{equation}
We aim to integrate over the high frequency $\p$
in the large-$N$ limit
and incorporate its effect as a renormalized
action of the low frequency $\bar{\Phi}$.
$\rho$ is then to play the r\^{o}le of a new cutoff.
Substituting $\P=\bar{\Phi}+\p$ into the action
(\ref{Svm}) it reads
\begin{eqnarray}
S[\P]&=&S[\bar{\Phi}]+\sigma[\bar{\Phi},\p]+S[\p],\nonumber\\
\sigma[\bar{\Phi},\p]&=&\frac{u}{N} \int {d^d x}\left[
2\bar{\Phi}^2\phi^2+4\bar{\Phi}^2\left(\bar{\Phi}\cdot\p\right)
+4\left(\bar{\Phi}\cdot\p\right)\phi^2+
4\left(\bar{\Phi}\cdot\p\right)^2 \right].
\label{sigma}
\end{eqnarray}
The quadratic term separates into $S[\bar{\Phi}]$ and $S[\p]$
due to momentum conservation.
Integration over $\p$ yields the induced action $\tilde{S}[\bar{\Phi}]$,
determined by
\begin{eqnarray}
\e^{-\tilde{S}[\bar{\Phi}]}&\equiv&
\left\langle \e^{-\sigma[\bar{\Phi},\p]} \right\rangle
=1-
\frac{u}{N} \int {d^d x} 
\left[ 2\bar{\Phi}^2(x)\left\langle \phi^2(x) \right\rangle
+\cdots\right]\nonumber\\
&&+\frac{1}{2} \left( \frac{u}{N} \right)^2 
\int\!\!\int{d^d x}\, {d^d y} \left[
4\bar{\Phi}^2(x)\left\langle \phi^2(x) \phi^2(y) \right\rangle
\bar{\Phi}^2(y)+\cdots \right]-\cdots.
\label{eSt}
\end{eqnarray}
Here $\left\langle \cdots \right\rangle$ denotes 
an average with respect to
the measure ${\cal D}\p\, e^{-S[\p]}$.

It is easy to confirm that the last three terms 
in eq.(\ref{sigma}) containing $(\bar{\Phi}\cdot\p)$ do not
contribute to $\tilde{S}$ in the large-$N$ limit.
Furthermore we truncate induced interactions to
those already present in the original action (\ref{Svm}).
This truncation to first three relevant terms is a natural
approximation for handling the RG transformation, 
although it is justified only a posteriori.
Then the terms exhibited explicitly in eq.(\ref{eSt}) suffice.
By reexponentiating the rhs of eq.(\ref{eSt}) we obtain
\begin{equation}
\tilde{S} [\bar{\Phi}]\!=\!
2\frac{u}{N} \int \bar{\Phi}^2(x)\left\langle \phi^2(x)
\right\rangle\!-\!
2 \left( \frac{u}{N} \right)^2 
\int\!\!\!\!\int \bar{\Phi}^2(x) \bar{\Phi}^2(y)
\left( \left\langle \phi^2(x) \phi^2(y) \right\rangle\!-
\!\left\langle \phi^2(x)
 \right\rangle\left\langle \phi^2(y) \right\rangle\right).
\label{Stvm}
\end{equation}

Now we are in a position to apply the Wilsonian
approximation to $\p$-correlators:
to replace all propagators $1/(p^2+r)$
appearing in the loop integrals with
$1/({\rm const.}+\nolinebreak r)$. 
Since the final result 
(eqs.(\ref{lmax},\ref{nuvm}))
is insensitive to the numerical value 
of the constant, we set it equal to unity.
This approximation virtually reduces correlators 
to zero-dimensional ones, which are exactly calculable
using the saddle point method \cite{HB}.
The (contracted) two-point function in eq.(\ref{Stvm}), 
\begin{eqnarray}
\left\langle \phi^2 \right\rangle&=&
\begin{picture}(100,15)(0,-3)
\put(0,7.5){\circle*{.5}}
\put(1.875,5.625){\circle*{.5}}
\put(3.75,3.75){\circle*{.5}}
\put(5.625,1.875){\circle*{.5}}
\put(7.5,0){\circle*{.5}}
\put(0,-7.5){\circle*{.5}}
\put(1.875,-5.625){\circle*{.5}}
\put(3.75,-3.75){\circle*{.5}}
\put(5.625,-1.875){\circle*{.5}}
\put(15,0){\circle{15}}
\put(30,-3){+}
\put(47.5,7.5){\circle*{.5}}
\put(49.375,5.625){\circle*{.5}}
\put(51.25,3.75){\circle*{.5}}
\put(53.125,1.875){\circle*{.5}}
\put(55,0){\circle*{.5}}
\put(47.5,-7.5){\circle*{.5}}
\put(49.375,-5.625){\circle*{.5}}
\put(51.25,-3.75){\circle*{.5}}
\put(53.125,-1.875){\circle*{.5}}
\put(62.5,0){\circle{15}}
\put(70,0){\circle*{2}}
\put(77.5,0){\circle{15}}
\put(92.5,-3){+\ $\cdots$}
\end{picture} \nonumber\\
&=&\int \frac{d^dp}{(2\pi)^d} \frac{N}{p^2+r}
 -4 \frac{u}{N} \int\frac{d^dp}{(2\pi)^d} \frac{N}{(p^2+r)^2}
\int \frac{d^dq}{(2\pi)^d} \frac{N}{q^2+r}+\cdots
\nonumber
\end{eqnarray}
(the real/dotted lines stand for $\p$/$\bar{\Phi}$ fields
respectively)
is approximated to be
\begin{equation}
\left\langle \phi^2 \right\rangle
\stackrel{{\rm approx.}}{\longrightarrow}
N\left( \frac{c_d}{1+r}-4\frac{u c_d^2}{(1+r)^3}+\cdots\right)
=N\frac{c_d}{1+r}
 C_2\left(\frac{u c_d}{(1+r)^2}\right) ,
\end{equation}
where $c_d=\int_{\rho < |p| \leq 1} {{d^d p}\over{(2\pi)^d}}$ 
and $C_2$ denotes the 
zero-dimensional two-point function in eq.(\ref{C2}).
Similarly the connected four-point function
at zero-external momenta
is approximated using
its zero-dimensional counterpart $C_4$ in eq.(\ref{C4}), 
\begin{eqnarray}
&&\left\langle \phi^2 \phi^2 \right\rangle
-\left\langle \phi^2 \right\rangle\left\langle \phi^2 \right\rangle=\;
\begin{picture}(150,22.5)(0,-3)
\put(0,7.5){\circle*{.5}}
\put(1.875,5.625){\circle*{.5}}
\put(3.75,3.75){\circle*{.5}}
\put(5.625,1.875){\circle*{.5}}
\put(7.5,0){\circle*{.5}}
\put(0,-7.5){\circle*{.5}}
\put(1.875,-5.625){\circle*{.5}}
\put(3.75,-3.75){\circle*{.5}}
\put(5.625,-1.875){\circle*{.5}}
\put(15,0){\circle{15}}
\put(22.5, 0){\circle*{.5}}
\put(28.125, 5.625){\circle*{.5}}
\put(26.25, 3.75){\circle*{.5}}
\put(24.375, 1.875){\circle*{.5}}
\put(30, 7.5){\circle*{.5}}
\put(28.125, -5.625){\circle*{.5}}
\put(26.25, -3.75){\circle*{.5}}
\put(24.375, -1.875){\circle*{.5}}
\put(30, -7.5){\circle*{.5}}
\put(38,-3){+}
\put(56,7.5){\circle*{.5}}
\put(57.875,5.625){\circle*{.5}}
\put(59.75,3.75){\circle*{.5}}
\put(61.625,1.875){\circle*{.5}}
\put(63.5,0){\circle*{.5}}
\put(56,-7.5){\circle*{.5}}
\put(57.875,-5.625){\circle*{.5}}
\put(59.75,-3.75){\circle*{.5}}
\put(61.625,-1.875){\circle*{.5}}
\put(71,0){\circle{15}}
\put(78.5,0){\circle*{2}}
\put(86,0){\circle{15}}
\put(101, 7.5){\circle*{.5}}
\put(99.125, 5.625){\circle*{.5}}
\put(97.25, 3.75){\circle*{.5}}
\put(95.375, 1.875){\circle*{.5}}
\put(93.5, 0){\circle*{.5}}
\put(101, -7.5){\circle*{.5}}
\put(99.125, -5.625){\circle*{.5}}
\put(97.25, -3.75){\circle*{.5}}
\put(95.375, -1.875){\circle*{.5}}
\put(109,-3){+}
\put(127,7.5){\circle*{.5}}
\put(128.875,5.625){\circle*{.5}}
\put(130.75,3.75){\circle*{.5}}
\put(132.625,1.875){\circle*{.5}}
\put(134.5,0){\circle*{.5}}
\put(127,-7.5){\circle*{.5}}
\put(128.875,-5.625){\circle*{.5}}
\put(130.75,-3.75){\circle*{.5}}
\put(132.625,-1.875){\circle*{.5}}
\put(142,7.5){\circle*{2}}
\put(142,0){\circle{15}}
\put(142,15){\circle{15}}
\put(157, 7.5){\circle*{.5}}
\put(155.125, 5.625){\circle*{.5}}
\put(153.25, 3.75){\circle*{.5}}
\put(151.375, 1.875){\circle*{.5}}
\put(149.5, 0){\circle*{.5}}
\put(157, -7.5){\circle*{.5}}
\put(155.125, -5.625){\circle*{.5}}
\put(153.25, -3.75){\circle*{.5}}
\put(151.375, -1.875){\circle*{.5}}
\put(165,-3){+\ $\cdots$}
\end{picture} \nonumber\\
&=&2\int \frac{d^dp}{(2\pi)^d} \frac{N}{(p^2+r)^2}
-8 \frac{u}{N}\int \frac{d^dp}{(2\pi)^d} \frac{N}{(p^2+r)^2}
 \int\frac{d^dq}{(2\pi)^d} \frac{N}{(q^2+r)^2}\nonumber\\
&&-16 \frac{u}{N}\int \frac{d^dp}{(2\pi)^d} \frac{N}{(p^2+r)^3}
 \int\frac{d^dq}{(2\pi)^d} \frac{N}{q^2+r}
+\cdots\nonumber\\
&\stackrel{{\rm approx.}}{\longrightarrow}&
N\left( \frac{2c_d}{(1+r)^2}-24\frac{u c_d^2}{(1+r)^4}
+\cdots\right)
=N\frac{c_d}{(1+r)^2} 
C_4\left( \frac{u c_d}{(1+r)^2}\right).
\end{eqnarray}
A characteristic feature of vector models is that
the momentum-dependent self-energy diagrams which could have
contributed to wave function renormalization vanish
in the large-$N$ limit. 
Thus the renormalized action for low frequency $\bar{\Phi}$ reads
\begin{eqnarray}
S'[\bar{\Phi}]&=&S[\bar{\Phi}]+\tilde{S}[\bar{\Phi}] 
\label{Sprime}\\
&=&\int{d^d x}\left[\frac{1}{2} (\nabla \bar{\Phi})^2+
\frac{1}{2} \left[ r+ 4 (1+r)g\, C_2(g) \right] \bar{\Phi}^2+
\frac{u}{N}\left[ 1-2g\, C_4(g) \right] 
\left( \bar{\Phi}^2\right)^2 \right]\nonumber
\end{eqnarray}
with $g=u\,c_d/(1+r)^2$.
Note that the coefficient functions of $\bar{\Phi}^2$ and 
$\left( \bar{\Phi}^2\right)^2$ by construction
have an interpretation as normalized 1PI vertices \cite{F},
\renewcommand{\theequation}{\arabic{equation}\alph{subeqn}}
\setcounter{subeqn}{1}
\begin{eqnarray}
r+ 4 (1+r)g\, C_2(g) &=& (1+r)\Gamma_2(g) -1,
\label{SDGa2}\\
\addtocounter{equation}{-1}
\addtocounter{subeqn}{1}
1-2g\, C_4(g)&=& \frac{\Gamma_4(g)}{8g}.
\label{SDGa4}
\end{eqnarray}
\renewcommand{\theequation}{\arabic{equation}}
They can be confirmed by Schwinger-Dyson equations\footnote{
I thank G.~Ferretti for pointing out this observation.};
eq.(\ref{SDGa4}) ($\times 8g$) is by
\begin{equation}
\begin{picture}(130,15)(15,-3)
\put(15,7.5){\circle*{.5}}
\put(16.875,5.625){\circle*{.5}}
\put(18.75,3.75){\circle*{.5}}
\put(20.625,1.875){\circle*{.5}}
\put(15,-7.5){\circle*{.5}}
\put(16.875,-5.625){\circle*{.5}}
\put(18.75,-3.75){\circle*{.5}}
\put(20.625,-1.875){\circle*{.5}}
\put(22.5, 0){\circle*{2}}
\put(28.125, 5.625){\circle*{.5}}
\put(26.25, 3.75){\circle*{.5}}
\put(24.375, 1.875){\circle*{.5}}
\put(30, 7.5){\circle*{.5}}
\put(28.125, -5.625){\circle*{.5}}
\put(26.25, -3.75){\circle*{.5}}
\put(24.375, -1.875){\circle*{.5}}
\put(30, -7.5){\circle*{.5}}
\put(38,-3){+}
\put(56,7.5){\circle*{.5}}
\put(57.875,5.625){\circle*{.5}}
\put(59.75,3.75){\circle*{.5}}
\put(61.625,1.875){\circle*{.5}}
\put(63.5,0){\circle*{2}}
\put(56,-7.5){\circle*{.5}}
\put(57.875,-5.625){\circle*{.5}}
\put(59.75,-3.75){\circle*{.5}}
\put(61.625,-1.875){\circle*{.5}}
\put(71,0){\oval(15,15)[l]}
\put(71,7.5){\line(1,0){15}}
\put(71,-7.5){\line(1,0){15}}
\put(78.5,2){\circle*{15}}
\put(78.5,-2){\circle*{15}}
\put(71,2){\line(0,-1){4}}
\put(86,2){\line(0,-1){4}}
\put(86,0){\oval(15,15)[r]}
\put(101, 7.5){\circle*{.5}}
\put(99.125, 5.625){\circle*{.5}}
\put(97.25, 3.75){\circle*{.5}}
\put(95.375, 1.875){\circle*{.5}}
\put(93.5, 0){\circle*{2}}
\put(101, -7.5){\circle*{.5}}
\put(99.125, -5.625){\circle*{.5}}
\put(97.25, -3.75){\circle*{.5}}
\put(95.375, -1.875){\circle*{.5}}
\put(109,-3){=}
\put(127,7.5){\circle*{.5}}
\put(128.875,5.625){\circle*{.5}}
\put(130.75,3.75){\circle*{.5}}
\put(132.625,1.875){\circle*{.5}}
\put(134.5,0){\circle*{.5}}
\put(127,-7.5){\circle*{.5}}
\put(128.875,-5.625){\circle*{.5}}
\put(130.75,-3.75){\circle*{.5}}
\put(132.625,-1.875){\circle*{.5}}
\put(142,0){\circle{15}}
\put(137.3,-2.65){$\times$}
\put(157, 7.5){\circle*{.5}}
\put(155.125, 5.625){\circle*{.5}}
\put(153.25, 3.75){\circle*{.5}}
\put(151.375, 1.875){\circle*{.5}}
\put(149.5, 0){\circle*{.5}}
\put(157, -7.5){\circle*{.5}}
\put(155.125, -5.625){\circle*{.5}}
\put(153.25, -3.75){\circle*{.5}}
\put(151.375, -1.875){\circle*{.5}}
\end{picture}
\end{equation}
(the solid/crossed blobs represent Green/1PI-vertex functions
$C_4$/$\Gamma_4$ respectively) and eq.(\ref{SDGa2}) is by eq.(\ref{sde}).

Finally we must rescale $x\rightarrow\rho^{-1}x$ ($p\rightarrow \rho\,p$) and
$\bar{\Phi}\rightarrow\rho^{d/2-1} \Phi$
so that the renormalized action (\ref{Sprime}) 
has the same momentum range $0\leq p \leq 1$
as the original one \nolinebreak(\ref{Svm}),
\begin{equation}
S'[\Phi]=
 \int {d^d x} \left[ \frac{1}{2}(\nabla \P)^2+
\frac{\rho^{-2}}{2} \left[ r+ 4 (1+r)g\, C_2(g) \right] 
\Phi^2+\frac{u}{N} 
\rho^{d-4} \left[ 1-2g \, C_4(g) \right] 
\left(\Phi^2\right)^2 \right] .
\end{equation}
Therefore the RG recursion equation takes the form ($\epsilon\equiv 4-d$)
\renewcommand{\theequation}{\arabic{equation}\alph{subeqn}}
\setcounter{subeqn}{1}
\begin{eqnarray}
r'=&\rho^{-2} \left[ r+ 4 (1+r)g\, C_2(g) \right]&
=\rho^{-2}  \left[ r+ (1+r)
{{-1 + {{(1 + 16\,g)^{1/2}}}}\over 2} \right] 
\label{r1vm} \\
\addtocounter{equation}{-1}
\addtocounter{subeqn}{1}
u'=& {u}\,\rho^{-\epsilon} \left[ 1-2g \, C_4(g) \right] &
=u\, {\rho^{-\epsilon}} { 1 + {{{{(1 + 16\,g)^{-1/2}}}}}\over 2}.
\label{u1vm}
\end{eqnarray}
\renewcommand{\theequation}{\arabic{equation}}
Now we are ready to solve the RG equation following
the general scheme. 
The non-gaussian fixed point is determined by $u'=u\neq 0$ 
in eq.(\ref{u1vm}), that is
\begin{equation}
{\rho^{\epsilon}}= { 1 + {{{{(1 + 16\,g_*)^{-1/2}}}}}\over 2}.
\label{fpc-g}
\end{equation}
For $\epsilon\ll 1$ eq.(\ref{fpc-g}) always has a solution
$g_*\sim\epsilon/4\,\log (1/\rho)>0$ as it should.
For $\epsilon \simeq 1$, however, it ceases to have a solution
on the perturbative sheet $(1 + 16\,g)^{1/2}=+\sqrt{1 + 16\,g}$
for a sufficiently small $\rho$ because its rhs always exceeds $1/2$.
Since whether $\rho{>\atop<} 2^{-1/\epsilon}$ has no physical significance
we are obliged to continue the fixed point to the second sheet
$(1 + 16\,g)^{1/2}=-\sqrt{1 + 16\,g}$.
Consequently the fixed point moves from $g=0$ to $g=+\infty$ on the 
first sheet and then turns back to $g=0$ on the second sheet as
$\rho$ decreases from 1 to 0. 
Under this agreement the fixed point $r=r'=r_*$, 
$u=u'=u_*$, $g_*=u_*\,c_d/(1+r_*)^2$
in eq.(12) is given by
\setcounter{subeqn}{1}
\renewcommand{\theequation}{\arabic{equation}\alph{subeqn}}
\begin{eqnarray}
g_*&=& {{{{\rho }^\epsilon}\,\left( 1 - {{\rho }^\epsilon} \right) }\over 
{4\,{{\left( 1 - 2\,{{\rho }^\epsilon} \right) }^2}}},
\label{gcvm} \\
\addtocounter{equation}{-1}
\addtocounter{subeqn}{1}
r_*&=&
-{{{{\rho }^\epsilon}}\over {1 - {{\rho }^2} - {{\rho }^\epsilon} 
+ 2\,{{\rho }^{2 + \epsilon}}}}<0,
\label{rcvm} \\
\addtocounter{equation}{-1}
\addtocounter{subeqn}{1}
u_*&=&{
{{\rho }^\epsilon}\,
\left( 1 - {{\rho }^\epsilon} \right)\,
\left( 1 - \rho^2  \right)^2
\over 
{4\,c_d\,{{\left( 1 - {{\rho }^2} - {{\rho }^\epsilon} + 
2\,{{\rho }^{2 + \epsilon}} \right) }^2}}} >0. 
\label{ucvm}
\end{eqnarray}
\renewcommand{\theequation}{\arabic{equation}}
The signs of $r_*$ and $u_*$ are in accord with
the general feature of the Wilson-Fisher fixed point
for $2<d<4$.
In order to calculate critical exponents we need to
linearize the RG equation (12) around $(r_*,\, u_*)$,
\begin{equation}
\left(
\begin{array}{c}
r'-r_* \\ u'-u_* 
\end{array} 
\right)
=
\left(
\matrix{ {{\rho }^{-2 + \epsilon}} & {{{{\rho }^\epsilon}\,
\left( -1 + {{\rho }^2} \right) \,
{{\left( -1 + {{\rho }^\epsilon} \right) }^2}}\over 
{2\,c_d\,\left( 1 - {{\rho }^2} - {{\rho }^\epsilon} +
2\,{{\rho }^{2 + \epsilon}} \right) }} \cr 
{{4\,c_d\,\left( 1 - {{\rho }^2} - {{\rho }^\epsilon} 
+ 2\,{{\rho }^{2 + \epsilon}} \right) }\over 
\rho^2\, (-1+\rho^2) } & 2 - 3\,{{\rho }^\epsilon} 
+ 2\,{{\rho }^{2\,\epsilon}} \cr  }
\right)
\left(
\begin{array}{c}
r-r_* \\ u-u_*
\end{array}
\right)  .
\end{equation}
The eigenvalues of the matrix above, 
\begin{equation} 
\lambda_{1, 2}=
1+ {{{{\rho }^{\epsilon - 2}}}/ 2} - {{3\,{{\rho }^\epsilon}}/ 2} 
+ {{\rho }^{2\,\epsilon}} \pm {\sqrt{
{{\left( 1 + {{{{\rho }^{\epsilon - 2}}}/ 2} - 
{{3\,{{\rho }^\epsilon}}/ 2} + {{\rho }^{2\,\epsilon}} \right)^2}}
-{{\rho }^{2\,\epsilon-2}}}} ,
\label{lmax}
\end{equation}
determines the first two of the series of scaling indices $y_m$
($m=1, 2, \cdots$) by \cite{W}
\begin{equation}
y_m=-\frac{\log \lambda_{m}}{\log \rho},
\end{equation}
the greatest of which is related to the mass exponent $\nu$
\begin{equation}
\nu=
\frac{1}{y_1}=
{1\over 2} + {\epsilon\over 4} + 
\left( {1\over 8} + {{{{\rho }^2}\,\log \rho }
\over {2\,\left( 1 - {{\rho }^2} \right) }}
\right) \epsilon^2+
\left( {1\over {16}} + {{{{\rho }^2}\,\log \rho }\over 
 {2\,\left( 1 - {{\rho }^2} \right) }}
\right) \epsilon^3+ O\left( \epsilon^4 \right).
\label{nuvm}
\end{equation}
Due to the approximation $1/(p^2+r)\rightarrow 1/(1+r)$, 
$O(\epsilon^2)$ or 
higher order terms depend on $\rho$, the portion of integrated momentum
region by one step and scaling is apparently broken.  
However we can confirm that $y_{1,2}$ are smooth 
under the switchover of the sheets 
at $\rho=2^{-1/\epsilon}$.
Moreover, in the $\rho\rightarrow 0$ limit\footnote{
Although $\rho\rightarrow 0$ corresponds to $g_*\rightarrow 0$, it
should not be confused with that the perturbative calculation suffices;
the fixed point is indeed on the second sheet of mapping
$g\mapsto C_{2,4}(g)$ and
this double-sheeted structure of Green functions is 
highly nonperturbative.}
the eigenvalues approach
$\lambda_{1}\rightarrow \rho^{\epsilon-2}$,
$\lambda_{2}\rightarrow \rho^{\epsilon}$
and provide us with the exact values for 
large-$N$ vector/spherical models in $2<d<4$,
$y_{m}=d-2m$, 
$\nu=1/(d-2)$ \cite{ZJ, WHM}.
We expect that $y_m$ for $m\geq 3$ be reproduced
by relaxing the truncation of induced interactions.
This exactness might be attributed to that 
in the limit $\rho\rightarrow 0$ 
the cutoff theory is so strongly
course-grained by a single step of RG transformation 
that it flows into the limiting IR theory quickly enough  
to exceed the accumulation of errors in the approximation. 
We will exploit this observation to calculate critical
exponents of matrix models in the subsequent section.\\

%%%%%%%%%%%%%% SECTION 3: MATRIX MODEL %%%%%%%%%%%%%%
\centerline{{\bf 3. Wilsonian approximated RG for matrix models}}

Application of the Wilsonian approximation to matrix models
was already considered in ref.\cite{F}; here the outline of 
derivation of the RG equation is briefly recalled.

We start from a Euclidean action of an 
$N\times N$ hermitian matrix field $\P(x)$,
\begin{equation}
S[\Phi]=\int {d^d x} \tr\left[
\frac{1}{2} (\nabla\Phi)^2 + \frac{r}{2} \Phi^2
+{u\over N} \Phi^4 \right]
\label{Smm}
\end{equation}
equipped with a cutoff=1 as before.
We separate $\P$ 
with respect to a momentum $\rho$
into low/high-frequency parts
$\P=\bar{\Phi}+\p$, whose coupling reads
\begin{equation}
\sigma[\bar{\Phi}, \phi] = 
{u\over N}\int{d^d x}\tr\bigg[4\bar{\Phi}^3\phi +
4\bar{\Phi}^2\phi^2 +
 2\bar{\Phi}\phi\bar{\Phi}\phi + 4\bar{\Phi}\phi^3\bigg].
\end{equation}
In the case of matrix models, $\p$-integration also induces 
products of traces such as $(\tr\bar{\Phi}^2)^2$
which is as relevant as the single trace, $\tr\bar{\Phi}^4$.
However we can still truncate induced interactions to those present 
in the action (\ref{Smm}) consistently because the
$(\tr\bar{\Phi}^2)^2$ term is induced always with a suppression factor
$1/N^2$ relative to $\tr\bar{\Phi}^4$
and thus negligible in the
$N\rightarrow\infty$ limit. Taking into account 
planarity of large-$N$ matrix models, ${\bf Z}_2$ symmetry
$\p\leftrightarrow -\p$ and momentum conservation,
the induced action reads
\begin{eqnarray}
 \tilde S[\bar{\Phi}] &=& 4{{u}\over{N}}\int
\left\langle\tr\bar{\Phi}^2\phi^2(x) \right\rangle\nonumber\\
&-& 8\left({{u}\over{N}}\right)^2\int\!\!\!\int
\Bigg( \left\langle\tr\bar{\Phi}\phi^3(x)\tr\bar{\Phi}\phi^3(y)
\right\rangle_{\rm conn.}
 + \left\langle\tr\bar{\Phi}^2\phi^2(x)\tr\bar{\Phi}^2\phi^2(y)
\right\rangle_{\rm conn.}
\Bigg)\nonumber\\
&+& 32\left({{u}\over{N}}\right)^3 \int\!\!\!\int\!\!\!\int
\left\langle\tr\bar{\Phi}^2\phi^2(x)\tr\bar{\Phi}\phi^3(y)
\tr\bar{\Phi}\phi^3(z)
\right\rangle_{\rm conn.}\nonumber\\
&-& \frac{32}{3}\left({{u}\over{N}}\right)^4 
\int\!\!\!\int\!\!\!\int\!\!\!\int
\left\langle\tr\bar{\Phi}\phi^3(x)\tr\bar{\Phi}\phi^3(y)
\tr\bar{\Phi}\phi^3(z)\tr\bar{\Phi}\phi^3(w)\right\rangle_{\rm conn.}. 
\label{Stmm}
\end{eqnarray}
Again we replace all propagators $1/(p^2+r)$ by
$1/(1+r)$ to approximate mass- and coupling constant
renormalization. The first two terms in eq.(\ref{Stmm}) 
contribute to the $\tr \bar{\Phi}^2$ term. By making use of
the SD equation, diagrammatically written as 
(the solid/crossed blobs represent 
Green/1PI-vertex functions $C_2$/$\Gamma_{2,4}$, respectively)
\pagebreak[3]
\begin{equation}
2\ \ \;
\begin{picture}(40,32)(0,0)
\put(0,0){\circle*{.5}}
\put(4,0){\circle*{.5}}
\put(8,0){\circle*{.5}}
\put(12,0){\circle*{.5}}
\put(16,0){\circle*{.5}}
\put(20,0){\circle*{.5}}
\put(24,0){\circle*{.5}}
\put(28,0){\circle*{.5}}
\put(32,0){\circle*{.5}}
\put(36,0){\circle*{.5}}
\put(0,4){\circle*{.5}}
\put(4,4){\circle*{.5}}
\put(8,4){\circle*{.5}}
\put(12,4){\circle*{.5}}
\put(16,4){\circle*{.5}}
\put(20,4){\circle*{.5}}
\put(24,4){\circle*{.5}}
\put(28,4){\circle*{.5}}
\put(32,4){\circle*{.5}}
\put(36,4){\circle*{.5}}
\put(16,16){\oval(24,24)[l]}
\put(16,16){\oval(16,16)[l]}
\put(20,16){\oval(24,24)[r]}
\put(20,16){\oval(16,16)[r]}
\put(16,8){\line(1,0){4}}
\put(18,26){\circle*{12}}
\end{picture} 
\ \ + \ \ \;
\begin{picture}(64,32)(0,-2.5)
\put(0,2){\circle*{.5}}
\put(4,2){\circle*{.5}}
\put(8,2){\circle*{.5}}
\put(12,2){\circle*{.5}}
\put(52,2){\circle*{.5}}
\put(56,2){\circle*{.5}}
\put(60,2){\circle*{.5}}
\put(64,2){\circle*{.5}}
\put(28,6){\oval(24,24)[tl]}
\put(28,6){\oval(32,32)[tl]}
\put(32,6){\oval(24,24)[tr]}
\put(32,6){\oval(32,32)[tr]}
\put(16,2){\line(1,0){24}}
\put(12,2){\line(0,1){4}}
\put(16,2){\line(0,1){4}}
\put(0,-2){\circle*{.5}}
\put(4,-2){\circle*{.5}}
\put(8,-2){\circle*{.5}}
\put(12,-2){\circle*{.5}}
\put(52,-2){\circle*{.5}}
\put(56,-2){\circle*{.5}}
\put(60,-2){\circle*{.5}}
\put(64,-2){\circle*{.5}}
\put(28,-6){\oval(24,24)[bl]}
\put(28,-6){\oval(32,32)[bl]}
\put(32,-6){\oval(24,24)[br]}
\put(32,-6){\oval(32,32)[br]}
\put(16,-2){\line(1,0){24}}
\put(12,-2){\line(0,-1){4}}
\put(16,-2){\line(0,-1){4}}
\put(30,0){\circle*{12}}
\put(30,20){\circle*{12}}
\put(30,-20){\circle*{12}}
\put(46,0){\circle{12}}
\put(41.5,-3){$\times$}
\end{picture}
\ \ =\ \ 
\begin{picture}(36,25)(0,-2.5)
\put(0,2){\circle*{.5}}
\put(4,2){\circle*{.5}}
\put(8,2){\circle*{.5}}
\put(12,2){\circle*{.5}}
\put(24,2){\circle*{.5}}
\put(28,2){\circle*{.5}}
\put(32,2){\circle*{.5}}
\put(36,2){\circle*{.5}}
\put(0,-2){\circle*{.5}}
\put(4,-2){\circle*{.5}}
\put(8,-2){\circle*{.5}}
\put(12,-2){\circle*{.5}}
\put(24,-2){\circle*{.5}}
\put(28,-2){\circle*{.5}}
\put(32,-2){\circle*{.5}}
\put(36,-2){\circle*{.5}}
\put(18,0){\circle{12}}
\put(13.5,-3){$\times$}
\end{picture}
\ \ -\ \ 
\begin{picture}(36,25)(0,-2.5)
\put(0,2){\circle*{.5}}
\put(4,2){\circle*{.5}}
\put(8,2){\circle*{.5}}
\put(12,2){\circle*{.5}}
\put(16,2){\circle*{.5}}
\put(20,2){\circle*{.5}}
\put(24,2){\circle*{.5}}
\put(0,-2){\circle*{.5}}
\put(4,-2){\circle*{.5}}
\put(8,-2){\circle*{.5}}
\put(12,-2){\circle*{.5}}
\put(16,-2){\circle*{.5}}
\put(20,-2){\circle*{.5}}
\put(24,-2){\circle*{.5}}
\end{picture}
\label{SD}
\end{equation}
\vspace{1pt}

\noindent
their contribution are summarized into
$\left( g={u\,c_d / (1+r)^2}\right)$
\begin{equation}
\tilde{S}_r[\bar{\Phi}]=(1+r)\frac{\Gamma_2(g)-1}{2}\int {d^d x} 
\tr \bar{\Phi}^2.
\end{equation}
Similarly, the induced interaction terms (the last three
in eq.(\ref{Stmm})) are neatly compiled into
\begin{equation}
\tilde{S}_u[\bar{\Phi}]=\frac{u}{N} 
\left(\frac{\Gamma_4(g)}{4g}-1 \right)\int{d^d x}\tr\bar{\Phi}^4
\end{equation}
under our approximation.

In the case of large-$N$ matrix models,
the second term in the lhs of eq.(\ref{SD})
contributes also to wave function renormalization.
To incorporate its contribution we need to 
differentiate it by the external momentum $p^2$ 
at $p=0$, which can not be treated in the original
ultra-local approximation. 
Following Golner's modification \cite{G}
justified on the dimensional ground,
we approximate this procedure simply by replacing
with multiplication of a propagator
\begin{equation}
\left.\frac{d}{dp^2}\right|_{p=0} 
\stackrel{{\rm approx.}}{\longrightarrow}
-\frac{1}{1+r}.
\end{equation}
Then the induced kinetic term is evaluated as
\begin{equation}
\tilde{S}_{p^2}[\bar{\Phi}]=2g\;\Gamma_4(g)C_2(g)^3
\int {d^d x} \tr \left( \nabla \bar{\Phi} \right)^2.
\end{equation}
To recapitulate, the renormalized action for low-frequency
$\bar{\Phi}$ reads
\begin{eqnarray}
S'[\bar{\Phi}]&=&S[\bar{\Phi}]+\tilde{S}[\bar{\Phi}]
=\int {d^d x} \tr \Bigg[ \frac{1}{2} \left[
1+4g\;\Gamma_4(g) C_2(g)^3 \right]
(\nabla \bar{\Phi})^2 \nonumber \\
&&+\frac{1}{2} \left[ (1+r)\Gamma_2(g) -1 \right] \bar{\Phi}^2+
\frac{u}{N} \frac{\Gamma_4(g)}{4g} 
\bar{\Phi}^4 \Bigg] . 
\label{Sprimemm}
\end{eqnarray}
After rescaling the kinetic term to a standard form and
then $x\rightarrow\rho^{-1}x$, 
$\bar{\Phi}\rightarrow\rho^{d/2-1} \Phi$, the
RG recursion equation takes the form
\setcounter{subeqn}{1}
\renewcommand{\theequation}{\arabic{equation}\alph{subeqn}}
\begin{eqnarray}
r'&=&\rho^{-2} \frac{(1+r)\Gamma_2(g) -1}{
1+4g\;\Gamma_4(g)C_2(g)^3}, 
\label{rprimemm} \\
\addtocounter{equation}{-1}
\addtocounter{subeqn}{1}
u'&=&u\, \rho^{-\epsilon} \frac{\Gamma_4(g)/4g}{
\left[ 1+4g\;\Gamma_4(g)C_2(g)^3 \right]^2} . 
\label{uprimemm}
\end{eqnarray}
\renewcommand{\theequation}{\arabic{equation}}

We are now ready to solve the RG eq.~utilizing
zero-dimensional Green functions \cite{BIPZ},
\begin{equation}
\Gamma_2=\frac{1}{C_2}= 
\frac{3}{a^2 (4-a^2)},\ \ \ 
\Gamma_4=-\frac{C_4}{(C_2)^4}=
\frac{9(1-a^2)(5-2a^2)}{a^4(4-a^2)^4}
\end{equation}
with $12g\;a^4+a^2-1=0$.
For $0<\epsilon<2$ there exists a unique non-gaussian
fixed point determined by eq.(\ref{uprimemm}),
\begin{equation}
\rho^{\epsilon}=\frac{\Gamma_4(g_*)/4g_*}{\left[ 1+4g_*\;
\Gamma_4(g_*)C_2(g_*)^3\right]^2}, 
\end{equation}
which turns out to move from $g_*=0$ to $\infty$
as $\rho=1\rightarrow 0$, always on the perturbative sheet
$a^2=(-1+\sqrt{1+48g})/(24g)$\footnote{
The wave function renormalization factor is responsible
for this fact; without it $g_*$ would proceed to the second sheet
as in the case of vector models and fail to possess a
meaningful $\rho \rightarrow 0$ limit for critical exponents this time.}.
We can again confirm $r_*<0$ and $u_*>0$.
The $y_{1,2}$ indices are obtained by
following the same procedure as in the previous section.
For any $\rho$ they turn out to lie in the range $y_1>0>y_2$
(and are equal to the mean field values $y_1=2,\ y_2=0$ for 
$\epsilon \rightarrow 0$ as should be),
in accordance with the fact that the Wilson-Fisher fixed point
for $2<d<4$ is associated with a single relevant perturbation.
The $O\left(\epsilon^2\right)$ and higher order terms
of the $\nu$ exponent depend on $\rho$ as before,
and converges to $\nu\rightarrow 1/(2-\epsilon/2)=2/d$ in the
$\rho\rightarrow 0$ limit (Fig.1).
\vskip -0.5cm
\epsfxsize=8cm
\centerline{\epsfbox{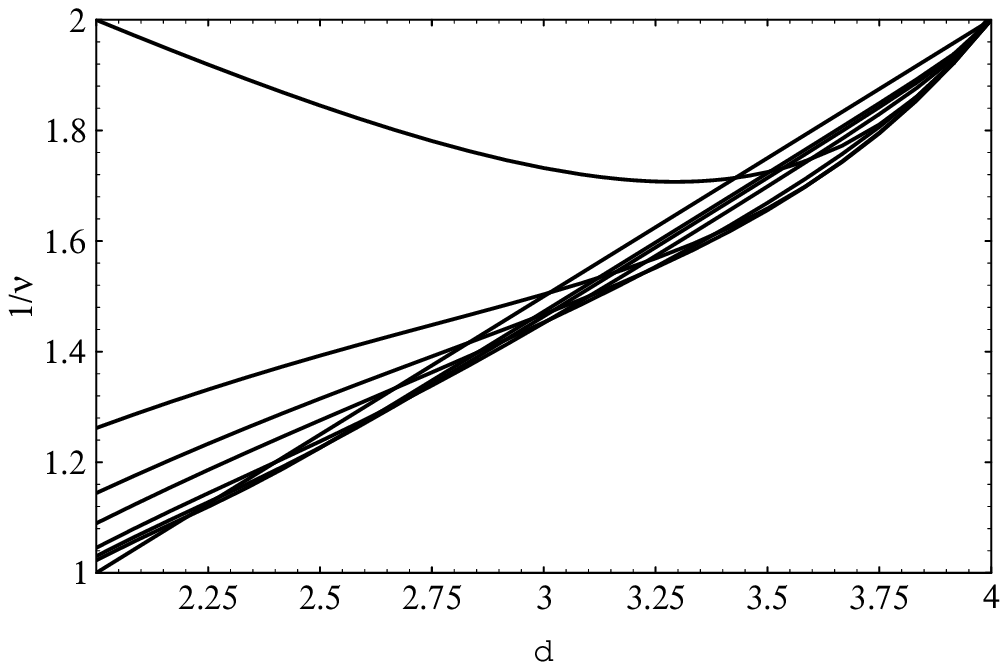}}
\vskip -1cm
\small
\centerline{Fig.1:\ Plot of $1/\nu$ for $2<d<4$.}
\centerline{
{}From top to bottom (at $d=2$): $\rho=1$, $1/2$, $1/4$, $10^{-1}$, 
$10^{-2}$, $10^{-3}$, $10^{-4}$ and 0.}
\normalsize
On the other hand, the anomalous dimension $\eta$, 
determined by
the wave function renormalization factor via
\begin{equation}
\eta=-\frac{\log \left[ 1+4g_*\;
\Gamma_4(g_*)C_2(g_*)^3 \right]}{\log \rho}
=\frac{\epsilon}{2} 
-\frac{\log (\Gamma_4(g_*)/4g_*)}{2\log \rho}
\end{equation}
can be shown to
converge to $\epsilon/2=2-d/2$ in the $\rho\rightarrow 0$ limit (Fig.2).
\vskip -0.5cm
\epsfxsize=8cm
\centerline{\epsfbox{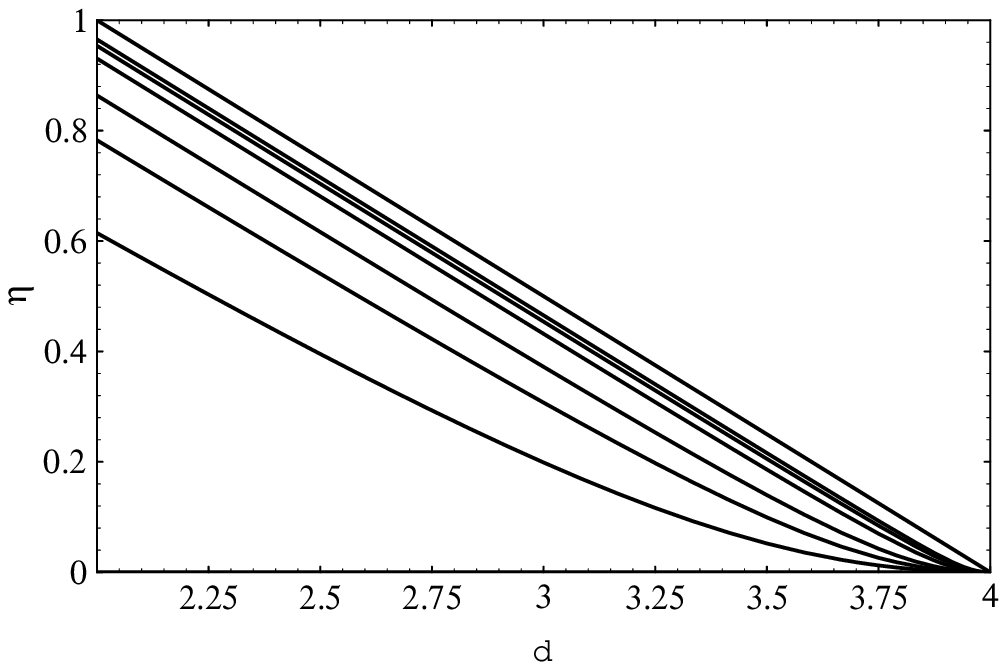}}
\vskip -1cm
\small
\centerline{Fig.2:\  Plot of $\eta$ for $2<d<4$.}
\centerline{
{}From bottom to top: $\rho=1$ ($\eta\equiv 0$), $1/2$, $1/4$, $10^{-1}$, 
$10^{-2}$, $10^{-3}$, $10^{-4}$ and 0.}
\normalsize
\noindent
Although the exactness of the $\rho\rightarrow 0$ limit in the case of
vector models does not necessarily imply that in matrix models,
it nevertheless provides us with a strong supporting ground for
these limiting values of the critical exponents.\\

%%%%%%%%%%%%%%% SECTION 4: REMARKS %%%%%%%%%%%%%%%
\centerline{{\bf 4. Concluding remarks}}

In this letter I have identified the Wilson-Fisher fixed point
(in the stable $u>0$ region)
both for large-$N$ vector and matrix models using the
Wilson's scheme and computed critical exponents.
The essential approximations employed are three-term truncation
of induced interactions
and zero-dimensionalization of the propagators, combined
with the $\rho\rightarrow 0$ limit.
This program is, in a sense, complementary to the standard approach
\cite{WHM} using the gap equation
where the approximation $1/(p^2+r)\rightarrow 1/(1+r)$
is known to yield the full series of exact $y_m$ indices
for large-$N$ vector models
without ambiguity in $\rho$,
after taking {\em all} the induced interactions into account.
What is remarkable for matrix models is that 
the critical exponents for magnetization and specific heat,
derived from (non-mean-field) $\nu$ and $\eta$ 
via (hyper-)scaling relations,
are predicted to stay at the classical mean field values
$\alpha=0,\ \beta=1/2,\ \gamma=1,\ \delta=3$,
despite $d$ is below the upper critical dimension 4.
This consequence is nontrivial and may not be attributed
to the roughness of the approximation when we recall its
exactness for vector models.

Generalization to higher-order truncation and criticality
as well as to non-hermitian matrix models is
straightforward.
Direct calculation of various magnetic exponents will be made possible
by relaxing the ${\bf Z}_2$ symmetry $\phi\leftrightarrow -\phi$, 
and serve for the check of consistency.
I hope to discuss these points in a subsequent publication.
Application of our program to large-$N$ QCD utilizing its low-dimensional
exact solution is another interesting subject, 
although a special care is required
for a cutoff procedure in order to maintain gauge invariance.

Finally I list a few points yet to be clarified.
The precise mechanism for the three-term truncation 
in the $\rho\rightarrow 0$ limit to work out successfully
for vector models must be fully explained in order to justify
the matrix model results. 
Turning back to the original motivation,
the relationship of these field-theoretic (spacetime) exponents 
to geometrical (world-sheet) exponents $d_{{\rm H}}$, $\gamma_{{\rm str}}$
etc., measured numerically for $c>1$ candidates \cite{AJW},
is also unclear to the author at present.\\

\centerline{{\bf Acknowledgements}}

I thank G.~Ferretti for discussion and explanation
of his paper. 
Thanks are also due to J.~Ambj{\o}rn and P.~H.~Damgaard
for discussions,
D.~Smith for checking the manuscript, M.~Moshe and colleagues
of Department of Physics, Technion for hospitality during my stay.
This work is supported by JSPS Postdoctoral Fellowships
for Research Abroad.\\

%%%%%%%%%%%%%%%%%%% APPENDIX %%%%%%%%%%%%%%%%%%%
\renewcommand{\theequation}{A\arabic{equation}}
\setcounter{equation}{0}
\centerline{{\bf Appendix: Green functions of 0D vector model}}

Here I summarize the derivation of Green functions of 
zero-dimensional vector model following ref.\cite{HB}.
The partition function is
\begin{eqnarray}
Z&=&\int  d^N\p\ \e^{-[\frac{1}{2} \p^2+\frac{g}{N} (\p^2)^2]}
\nonumber\\
&=&\int_{-\infty}^\infty dt\ 
\e^{-N[\frac{1}{2} \e^t+g\ \e^{2t}-\frac{1}{2} t]}\equiv
\int dt\ \e^{-N\,F(t)} \nonumber\\
&=&\e^{-N\, F(t_s)-\frac{1}{2}\log F''(t_s)+O(1/N)}
\end{eqnarray}
where $\e^t\equiv\p^2/N$. The saddle point 
$t_s$ in the above is determined by
\begin{equation}
F'(t_s)=\frac{1}{2} \e^{t_s}+2g\ \e^{2t_s}-\frac{1}{2}=0.
\end{equation}
The four-point function is
\begin{equation}
\left\langle (\p^2)^2 \right\rangle=-N \frac1Z \frac{dZ}{dg}=
N^2\, \e^{2t_s}+N\, \frac{d}{dg}
 \left( \frac{1}{2} \log  F''(t_s)\right)+
O\left( 1 \right).
\end{equation}
On the other hand, making use of the SD equation
\begin{equation}
0=\frac{1}{Z}
\int d^N\p \sum_{i=1}^N \frac{d}{d\p_i}\left\{ \p_i
\ \e^{-[\frac{1}{2} \p^2+\frac{g}{N} (\p^2)^2]} \right\}
= N-\left\langle \p^2\right\rangle -4\frac{g}{N} 
\left\langle (\p^2)^2\right\rangle
\label{sde}
\end{equation}
the two-point function is given by
\begin{eqnarray}
\left\langle \p^2\right\rangle&=&N-4\frac{g}{N} 
\left\langle (\p^2)^2\right\rangle\nonumber\\
&=&N\,\e^{t_s}-4g \frac{d}{dg} \left( \frac{1}{2} \log F''(t_s)\right)
+O\left(\frac{1}{N}\right).
\end{eqnarray}
To recapitulate,
\begin{eqnarray}
C_2(g)&\equiv&\lim_{N\rightarrow\infty}\frac1N 
\left\langle \p^2\right\rangle=\e^{t_s}\nonumber\\
&=&{{-1 + {(1 + 16\,g)^{1/2}}}\over {8\,g}}
=1-4\,g+32\,g^2-\cdots 
\label{C2}\\
C_4(g)&\equiv&\lim_{N\rightarrow\infty}
\frac1N \left[ 
\left\langle (\p^2)^2 \right\rangle - 
\left\langle \p^2\right\rangle^2 \right]=
\left( 1+8g\, \e^{t_s}\right) \frac{d}{dg} 
\left( \frac{1}{2} \log F''(t_s)\right)\nonumber\\
&=&
\frac{{1 - {{{(1 + 16\,g)^{-1/2}}}}}}{4\, g} 
=2-24\,g+320\,g^2-\cdots .
\label{C4}
\end{eqnarray}

%%%%%%%%%%%%%%%%%% REFERENCES %%%%%%%%%%%%%%%%%%%

\end{document}